\documentclass[aps,showkeys]{revtex4}

\usepackage{epsfig}
\usepackage{eepic}
\usepackage{latexsym}
\usepackage{rotating}
\usepackage{lscape}
\usepackage{graphics}
\usepackage{dcolumn}
\usepackage{bm}
\newcommand{\beq}{\begin{equation}}
\newcommand{\eeq}{\end{equation}}
\newcommand{\bd}{\begin{displaymath}}
\newcommand{\ed}{\end{displaymath}}

\newcommand{\bei}{\begin{itemize}}
\newcommand{\eei}{\end{itemize}}

\newcommand{\bee}{\begin{enumerate}}
\newcommand{\eee}{\end{enumerate}}

\begin{document}

\noindent
\title{Monte Carlo study for designing a dedicated `D'-shaped collimator used in the external beam radiotherapy of retinoblastoma patients}

\author{P. A. Mayorga$^{1,2}$, L. Brualla$^3$, W. Sauerwein$^3$ and A. M. Lallena$^2$\\
{\small {\it 
$^1$Departamento de F\'{\i}sica, Facultad de Ciencias, Pontificia Universidad Javeriana, Cra. 7 40-62, Bogot\'a D C, Colombia. \\
$^2$Departamento de F\'{\i}sica At\'omica, Molecular y Nuclear, Universidad de Granada, E-18071 Granada, Spain \\ 
$^3$NCTeam, Strahlenklinik, Universit\"atsklinikum Essen, Hufelandstra\ss e 55, D-45122 Essen, Germany.
}}}

\date{\today}

\bigskip

\begin{abstract}
\noindent
{\it Purpose:} Retinoblastoma is the most common intraocular malignancy in the early childhood. Patients treated with external beam radiotherapy respond very well to the treatment. However, owing to the genotype of children suffering hereditary retinoblastoma, the risk of secondary radio-induced malignancies is high. The University Hospital of Essen has successfully treated these patients on a daily basis during nearly 30 years using a dedicated `D'-shaped collimator. The use of this collimator, that delivers a highly-conformed small radiation field, gives very good results in the control of the primary tumor as well as in preserving visual function, while it avoids the devastating side effects of deformation of midface bones. The purpose of the present article is to propose a modified version of the `D'-shaped collimator that reduces even further the irradiation field with the scope to reduce as well the risk of radio-induced secondary malignancies. Concurrently, the new dedicated `D'-shaped collimator must be easier to build and at the same time produce dose distributions that only differ on the field size with respect to the dose distributions obtained by the current collimator in use. The scope of the former requirement is to facilitate the employment of our irradiation technique both at our and at other hospitals. The fulfillment of the latter allows us to continue using the clinical experience gained in more than 30 years.\\
{\it Methods:} The Monte Carlo code {\sc penelope} was used to study the effect that the different structural elements of the dedicated `D'-shaped collimator have on the absorbed dose distribution. To perform this study the radiation transport through a Varian Clinac 2100~C/D operating at 6~MV was simulated in order to tally phase-space files which were then used as radiation sources to simulate the considered collimators and the subsequent dose distributions. With the knowledge gained in that study a new, simpler, `D'-shaped collimator is proposed.\\ 
{\it Results:} The proposed collimator delivers a dose distribution which is 2.4~cm wide along the inferior-superior direction of the eyeball. This width is 0.3~cm narrower than that of the dose distribution obtained with the collimator currently in clinical use. The other relevant characteristics of the dose distribution obtained with the new collimator, namely, depth doses at clinically relevant positions, penumbrae width and shape of the lateral profiles, are statistically compatible with the results obtained for the collimator currently in use.\\ 
{\it Conclusions:} The smaller field size delivered by the proposed collimator still fully covers the planning target volume with at least 95\% of the maximum dose at a depth of 2~cm and provides a safety margin of 0.2~cm, so ensuring an adequate treatment while reducing the dose absorbed by surrounding structures of the eye.
\end{abstract}


\maketitle

\section{Introduction}

Eye tumors require dedicated radiation techniques. These techniques are demanding owing to the close connection of the tumor to sensitive structures and also because of the small size of the target volume, which represents a challenge from the dosimetry point of view. Applicators using $^{106}$Ru/$^{106}$Rh \cite{Schueler06a} and external beam radiotherapy (EBRT) with either protons \cite{Mourtada,Woong}, electrons \cite{Brualla11b, Brualla09b} or photons \cite{Sauerwein} are among the established techniques. Common to all these modalities is the necessity of an accurate determination of the dose distribution inside the eye and in the surrounding structures and tissues \cite{Fluhs, ICRU, Schueler06b}. In this respect, Monte Carlo (MC) simulation has permitted a significant advance in the last years, providing an accurate description of the dose imparted to these small-sized irradiated volumes \cite{Chiu, Miras, Thomson08, Brualla12c}.

Retinoblastoma (Rb), the most common intraocular malignancy in the early childhood, responds very well to radiotherapy. Rb was treated with first line EBRT as the treatment of choice with a total dose from 40--50 Gy \cite{Messmer} resulting in a close to 90\% eye preservation in group I--IV (Reese-Ellsworth classification \cite{Reese}). There are two different types of Rb: a hereditary and a non-hereditary form. Both forms are induced by the biallelic inactivation of the RB1 tumor suppressor gene. The non-hereditary form results in a solitary tumor in one eye. For the hereditary variant the first mutation of the RB1 allele is inherited following the somatic inactivation of the second allele resulting in multiple tumor lesions in both eyes. The genotype in hereditary Rb patients predisposes the appearance of secondary malignancies, with radiotherapy further enhancing the risk of tumors occurring in the irradiated area. Treatments with EBRT have showed very good results in the control of primary tumors as well as in preserving the visual function \cite{Sauerwein}. However, during the last 15 years the trend has been to move away from radiotherapy mainly with the goal to reduce radiation induced secondary malignancies \cite{Abramson, Dommering, Turaka, Vasudevan, Rodjan}, but also in view of the devastating side effects originated by inappropriate irradiation techniques leading to deformation of the bones of the midface \cite{Imhof, Peylan}. This tendency has favored the use of other techniques in young children \cite{Gallie, Gobin, Munier, Schueler06a, Schueler06b, Shields96, Temming}. Most of them involve chemotherapy giving rise to damages in the retina, a reduction of the visual acuity, an increase of the toxicity on other organs \cite{Qaddoumi}, and an increase of the incidence of leukaemia and solid tumors \cite{Felix, Gombos}. In case of failure of these therapies, EBRT remains a salvage option, but adding further local damage and augmenting the risk of secondary malignancies.

In these circumstances EBRT can be reconsidered as first line treatment of Rb using dedicated treatment techniques that avoid asymmetric growth of the face while protecting sensitive structures of the eye. In this article we examine the possibility to further improve the lens sparing irradiation technique, which was initially proposed by Schipper \cite{Schipper83, Schipper97}, by reducing the irradiated volume and limiting further the penumbra.

In a previous work \cite{Brualla12a} we assessed by MC simulation the absorbed dose distribution in a water phantom obtained with the Shipper's technique \cite{Sauerwein09} as it is currently applied at the University Hospital of Essen. The technique consists of an EBRT irradiation that uses a dedicated `D'-shaped collimator whose purpose is to minimize the absorbed dose to normal tissues, hence reducing the incidence of radio-induced secondary tumors. This collimator can conform two field sizes depending whether an additional brass insert is employed or not. The large field (5.2~cm$^2$) is conformed without the brass insert and it is usually employed in children older than 6 months. For the small field (3.1~cm$^2$), employed in children younger than 6 months old, the brass insert is required. In the present article we investigate, by using the MC code {\sc penelope} \cite{PENELOPE}, the role played by the various pieces conforming the collimation system, paying attention to both the shape of its constructive elements and the materials conforming them. The ultimate goal is to propose a modified collimator that even reduces further the field size of the large field conformed without the brass insert, while maintaining the shape of the penumbrae and guaranteeing at least 95\% of the maximum dose along 2.4~cm of a centered lateral profile in the $x$ direction (corresponding to the inferior-superior direction of the eyeball) measured at a depth of 2~cm in a water phantom. The scope of reducing the field size is to pursue a lower incidence of radio-induced secondary tumors by reducing the volume of irradiated normal tissue in children older than 6 months. The aim of mantaining the characteristics of the dose distribution, while only reducing the field size, is to be able to continue applying the clinical experience gained in nearly 30 years of daily application of this irradiation technique at Essen. With respect to the collimator with the brass insert in use (small field), it was found that the irradiated volume is adequate for children younger than 6 months old and, therefore, no modification is proposed.

The studies conducting to the currently used version of the modified Shipper's technique were based on experiments and calculations done on water phantoms. Adequate computerized tomograpies of young children are rare and we do not have any. Thus, the actual patient's geometry is not available to MC treatment planning programs for its simulation. Treatment planning must resort to results obtained in water phantoms. For the design of the proposed collimator in this article we have compared our absorbed dose distributions in water phantoms, obtained with MC simulations, with results from the currently used collimator also obtained with MC simulations in water phantoms. 

About 60\% of the Rb patients irradiated at Essen have one eye enucleated at the time of treatment with EBRT. These patients are irradiated with a single field. Non-hereditary Rb patients with a single tumor in one eye are also irradiated with a single field. Only patients with tumors in both eyes at the time of radiotherapy are irradiated with two opposed fields. The modified collimator proposed herein has been optimized for a single field irradiation, so to serve to the majority of the treated Rb patients.

\section{Material and methods}

\subsection{`D'-shaped collimation system currently in use}

In this work we have studied the dosimetry of a `D'-shaped collimation system used for Rb treatment when various changes in the shape of its constructive elements and in the material composition of the collimator are considered. The Rb collimator is designed to be inserted in the accessory tray holder of a Varian Clinac~2100~C/D operating at 6~MV. Recently \cite{Brualla12a}, the dosimetry of this system was analyzed by comparing results obtained in MC simulations with {\sc penelope} \cite{PENELOPE}, experimental measurements in a water phantom and the predictions of the analytical anisotropic algorithm \cite{Sievinen, Ulmer95, Ulmer03} implemented in the Eclipse (Varian Medical Systems, Palo Alto, California) treatment planning system.

\begin{figure}[ht]
\centering
\includegraphics[width=8cm]{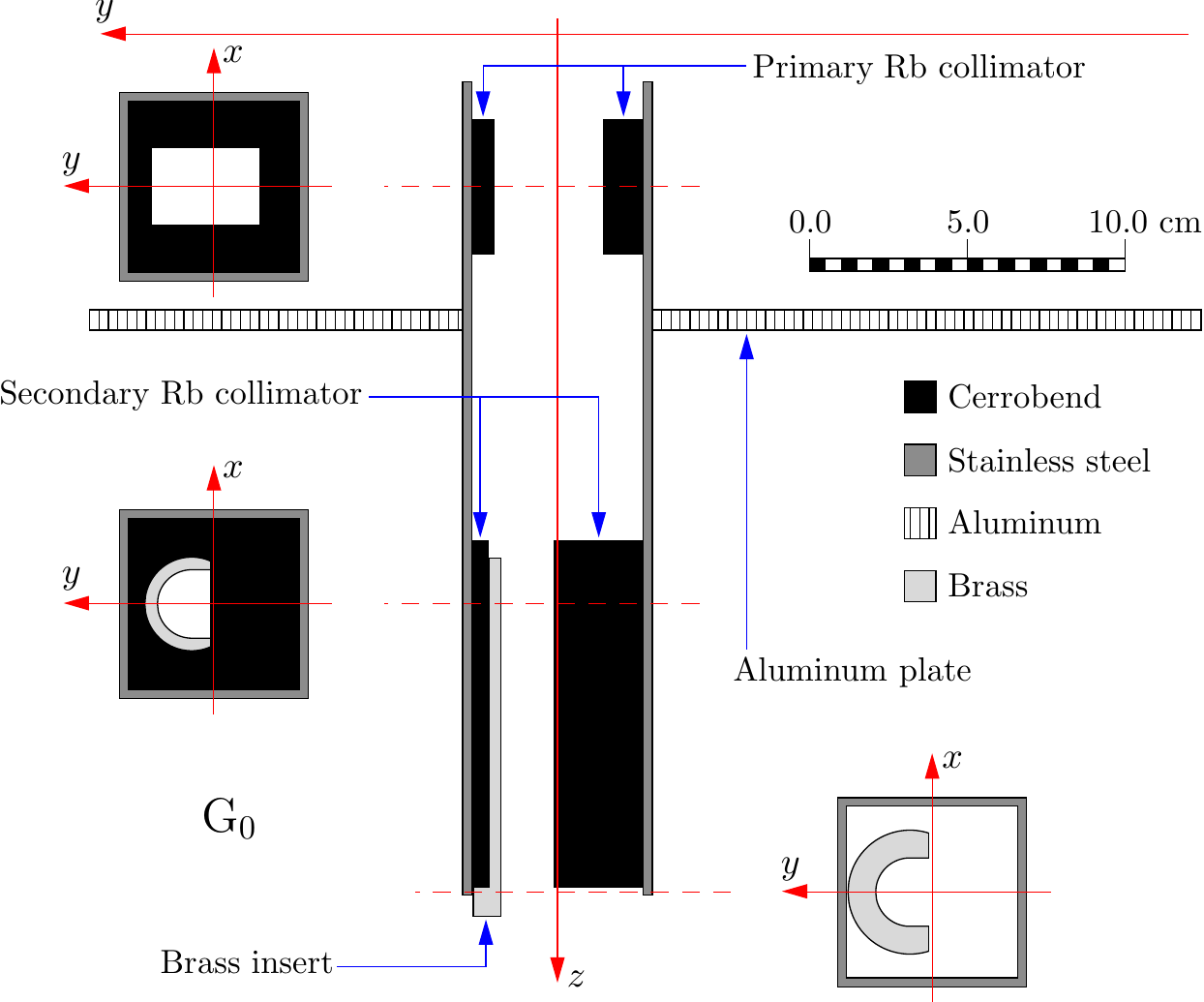}
\caption{Geometry G$_0$ corresponding to the dedicated `D'-shaped collimator studied in the present work. Dashed lines indicate the cuts at which the insets have been taken. Detailed blueprints of this collimator can be found in \cite{Brualla12a}.
\label{fig:G0}}
\end{figure}

Figure~\ref{fig:G0} shows the collimation system currently in use. It includes a primary and a secondary Rb collimator made of Cerrobend plus and optional brass insert that permits to reduce the radiation field maintaining its `D'-shape. Thus, patients can be treated with either $5.2$ cm$^2$ or $3.1$ cm$^2$ fields. Usually, patients younger than 6 months old are treated with the smaller field. The aluminum plate (figure~\ref{fig:G0}) permits to insert the collimator in the accessory tray holder of the linac. The beam central axis defines the direction of increasing $z$.

The coordinate system plotted in this figure is the one we have used for all calculations. The origin of coordinates is situated at $100$ cm from the isocenter upstream in the $z$ axis. The $x$-jaws of the linac define a symmetric field of $5.5$~cm width about the $y$ axis. The $y$-jaws in the negative and positive $y$-axis are displaced 0.7 and 3.5 cm, respectively, away from their closed position. A radiation field of $5.5\times4.2$~cm$^2$, centered at $x = 0$ and $y = 1.4$~cm is produced at the isocenter. The movable jaws are situated in the same place independently whether the brass insert is used or not. We called this geometry G$_0$ and we used it as reference to compare the results obtained with the modified geometries that we  studied. Results obtained with and without the brass insert were labeled as G$_0^{\rm w}$ and G$_0^{\rm wo}$, respectively. Depth doses at the center of the field, and $x$ and $y$ lateral profiles for $y = 1.3$~cm and $x = 0$, respectively, at three different depths ($z = 1.5$, $3.0$ and $5.0$~cm) were estimated in~\cite{Brualla12a}.

\begin{figure}[ht]
\centering
\includegraphics[width=8cm]{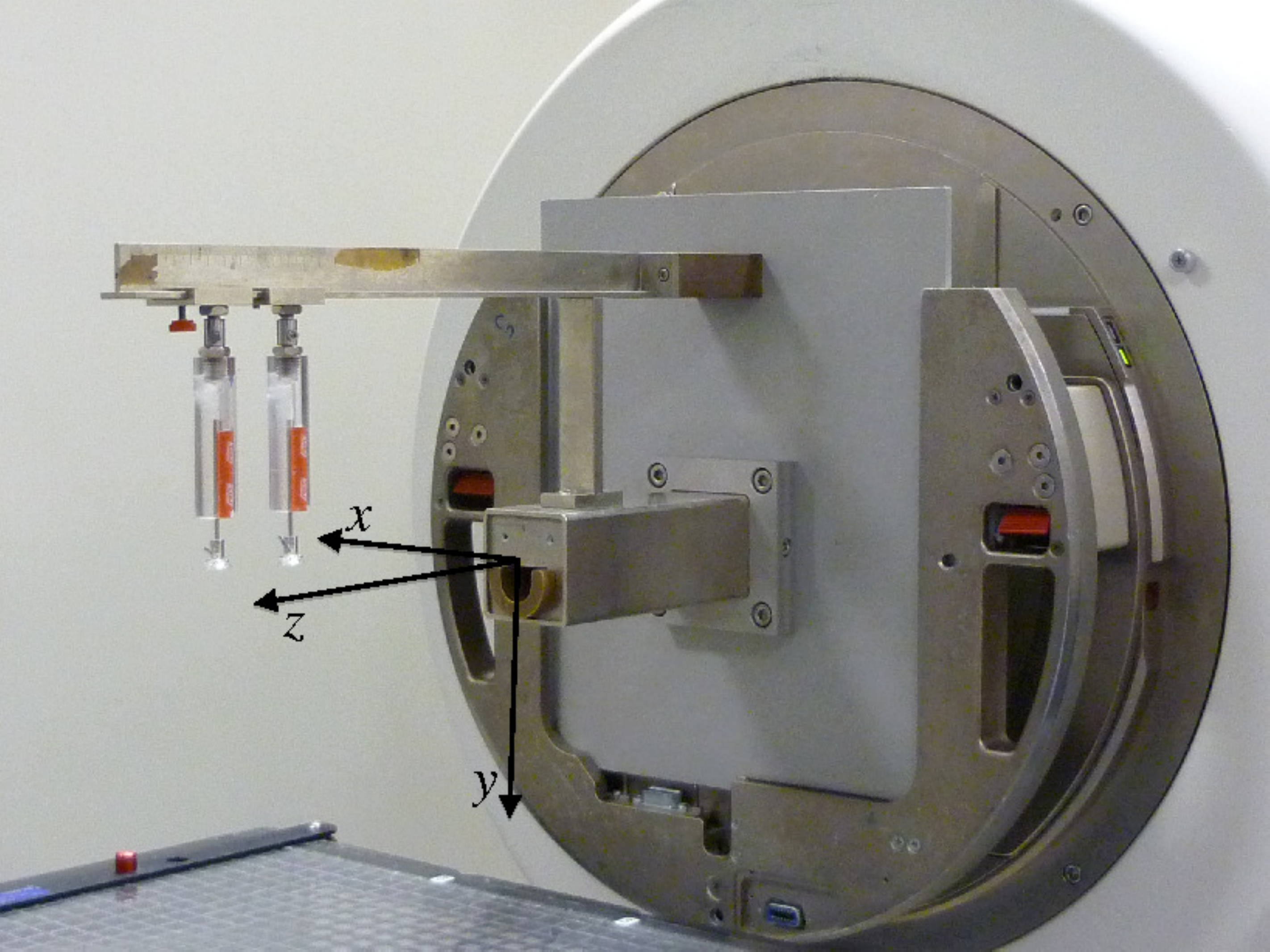}
\caption{Photograph of the Varian Clinac 2100~C/D set in position for irradiating a Rb patient. The patient is placed on the couch in head-first supine position, thus, his left eye is found upstream. The coordinate system used in the simulations presented herein is shown. The flat wall of the `D' is 1~mm away from the $z$-axis.
\label{fig:Picture}
}
\end{figure}

Figure~\ref{fig:Picture} shows a photograph of the Rb collimator mounted on a Varian Clinac~2100~C/D with its gantry and collimator angles set for irradiating a patient in head-first supine position (i.e., the left eye would be placed upstream). The scope of this figure and the following explanation is to relate the coordinate system used in the simulation with the anatomical directions of the eyeball. For clarity reasons the position of both eyes is related to the coordinate system, although, in the clinical practice most patients have one eye enucleated. The coordinate system defined in figures~\ref{fig:G0} and~\ref{fig:Picture}, relates to the anatomical directions of the eyeball as follows: i) the $x$-axis increases in the inferior-superior direction; ii) the $y$-axis increases in the anterior-posterior direction; iii) the $z$-axis increases in the lateral-nasal direction for the eye upstream and in the nasal-lateral direction for the eye downstream. During the treatment the eyes of the patient are fixed with the vacuum lenses appearing in the photograph mounted on the rail, so that the posterior pole of both eye lenses lie at $x=y=0$, that is, 1~mm away from the flat side of the `D' into the shielded part of the field. Thus, effectively protecting the eye lense and its surrounding structures from damaging radiation. 

\subsection{Simulation code}

All simulations were run using the main steering code {\sc penEasy}~\cite{Sempau11} that runs on the general-purpose MC system {\sc penelope} \cite{Baro, PENELOPE, Sempau97}. The geometry of the Varian Clinac~2100~C/D was automatically generated with the code {\sc penEasyLinac}~\cite{Brualla09b, Sempau11,Rodriguez2013}. The calculations with this set of codes were benchmarked specifically for the linac considered in this work operating at 6~MV (which is the nominal energy used in all simulations presented herein) in previous works~\cite{Brualla09b, Brualla09a, Fernandez, Panettieri, Sempau11}

{\sc penelope} simulates the coupled transport of electrons, positrons and photons and makes use of a mixed simulation scheme that classifies electron and positron interactions either as hard or soft events. On the one hand, hard events correspond to interactions in which the angular deflections and/or the energy losses are larger than certain cutoffs; these events are simulated in a detailed way. On the other hand, the soft interactions occurring between two hard events are simulated as a single artificial event described within a multiple scattering theory. The particles are transported while their kinetic energies are larger than the absorption energies defined by the user. In our case, these energies were set to 100~keV for electrons and positrons and 20~keV for photons. Electron and positron transport is controlled by five user defined parameters whose role is described in detail in the {\sc penelope} user's manual \cite{PENELOPE}. In our simulations {\sc penEasyLinac} set C1 = C2 = 0.1, WCC = 100~keV, WCR = 20~keV and DSMAX equals to one tenth of the thickness of the different geometry bodies along the beam path.

The depth dose and lateral dose profiles were scored in a $50 \times 50 \times 50$~cm$^3$ water phantom situated at a source-to-surface distance of 100~cm. The simulations were performed in three steps. In the first one, a phase-space file was tallied at the exit of the linac gantry for each considered configuration of the jaws. In this step the source was simulated as a monoenergetic point-like pencil beam with zero divergence and emitting electrons with an initial kinetic energy of 6.26~MeV. In the second step, the phase-space files scored in the first one were used as sources and particles were transported downstream through the corresponding `D'-shaped dedicated collimator with or without the brass insert. In this step, new phase-space files were tallied at the surface of the water phantom. Finally, absorbed dose distributions in water were obtained by transporting the particles in the phase-space files tallied in the second step into the water phantom. Absorbed dose distributions were tallied using scoring bins of $0.05\times0.05\times0.1$ cm$^3$.

\subsection{Modified geometries}

\begin{figure}[ht]
\centering
\includegraphics[width=8cm]{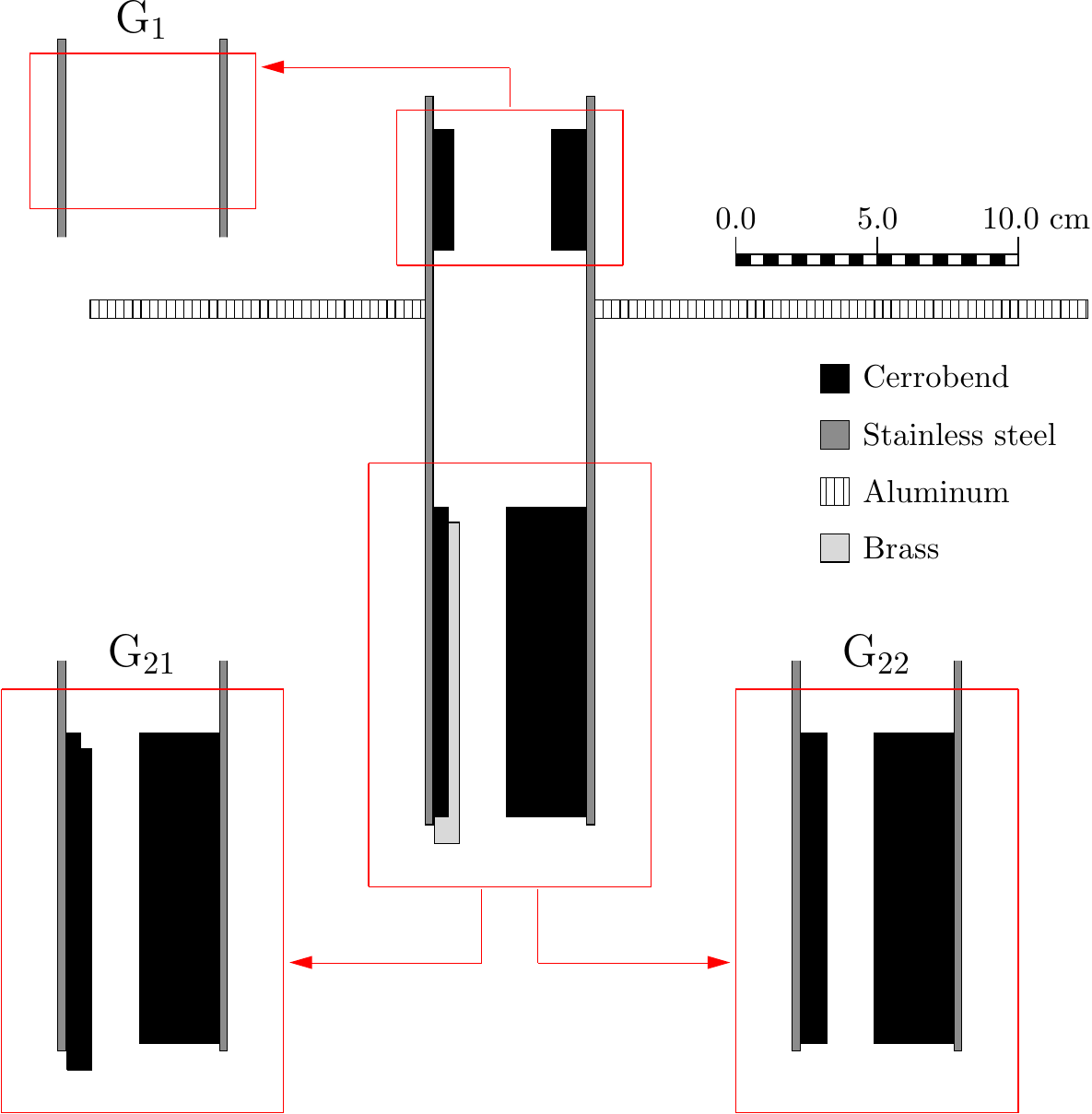}
\caption{Modified geometries studied in the present work.
\label{fig:Gmod}
}
\end{figure}

We are interested in investigating the dosimetric impact of various elements of the dedicated `D'-shape collimation system described in figure~\ref{fig:G0}. To that end, we studied a set of new geometries obtained by modifying the original geometry G$_0$. Three of these modifications are sketched in figure~\ref{fig:Gmod}.

With the first one, G$_1$, we have studied the role played by the primary collimator. As shown in figure~\ref{fig:Gmod}, in this geometry the primary collimator was eliminated, while the secondary collimator and the brass insert were kept inaltered. Simulations with and without the brass insert were performed and, as in the case of the G$_0$ geometry, we labeled the corresponding calculations as G$_1^{\rm w}$ and G$_1^{\rm wo}$, respectively.

The next group of geometries is related to the material and shape of the brass insert. First, we have assumed that this insert is made of Cerrobend (as the primary and the secondary Rb collimators are). The corresponding geometry was labeled G$_{21}$. In a second modification, the shape of the brass insert was changed to match exactly the length of the secondary collimator. We labeled G$_{22}$ this new geometry. The remaining elements of the collimator were maintained as in the G$_{0}$ geometry.

In these three modified geometries, the jaws were situated as described for the G$_{0}$ geometry. The radiation field defined by the jaws at the isocenter ($5.5\times 4.2$ cm$^2$, centered at $x = 0$ and $y = 1.4$ cm) is considerably larger than the field delimited by the Rb collimator which is, as said above, of 3.1 or $5.2$ cm$^2$ if the brass insert is included or not. In order to analyze how the jaw position can affect the final dose distribution, we considered a modified linac configuration in which the field delimited by the jaws was reduced to circumscribe the field defined by the Rb collimator, with and without the brass insert. Though the linac jaws are focalized to the target, the Rb collimator is not (i.e., its inner walls are parallel to the $z$-axis) and then two different positions of the jaws were analyzed where the circumscribed field corresponds to the fields delimited by the projections to the isocenter of the upstream and the downstream ends of the secondary Rb collimator. We labeled these geometries as G$_{\rm U}$ and G$_{\rm D}$, respectively. We simulated the geometries including or not the brass insert, without modifying the geometry of the Rb collimator.

The changes in the jaw positions produce a modification of the fluence and the dose delivered to the water phantom changes. To compare the dose distributions obtained with the modified geometries they were scaled so that all the maxima of the different dose distributions match that obtained with the G$_0$ geometry. The dose distributions obtained for the geometries G$_{\rm U}^{\rm wo}$, G$_{\rm U}^{\rm w}$, G$_{\rm D}^{\rm wo}$ and G$_{\rm D}^{\rm w}$ were multiplied by 1.010, 1.014, 1.018 and 1.023, respectively.

The MC standard statistical relative uncertainties for all bins with a dose larger than 50\% than the maximum dose was below 0.1\% in average. In the figures shown below, these uncertainties are smaller than the symbols used to represent the results obtained with the G$_0$ geometry.

\subsection{Profile analysis}

\begin{figure}[!b]
\centering
\includegraphics[width=8cm]{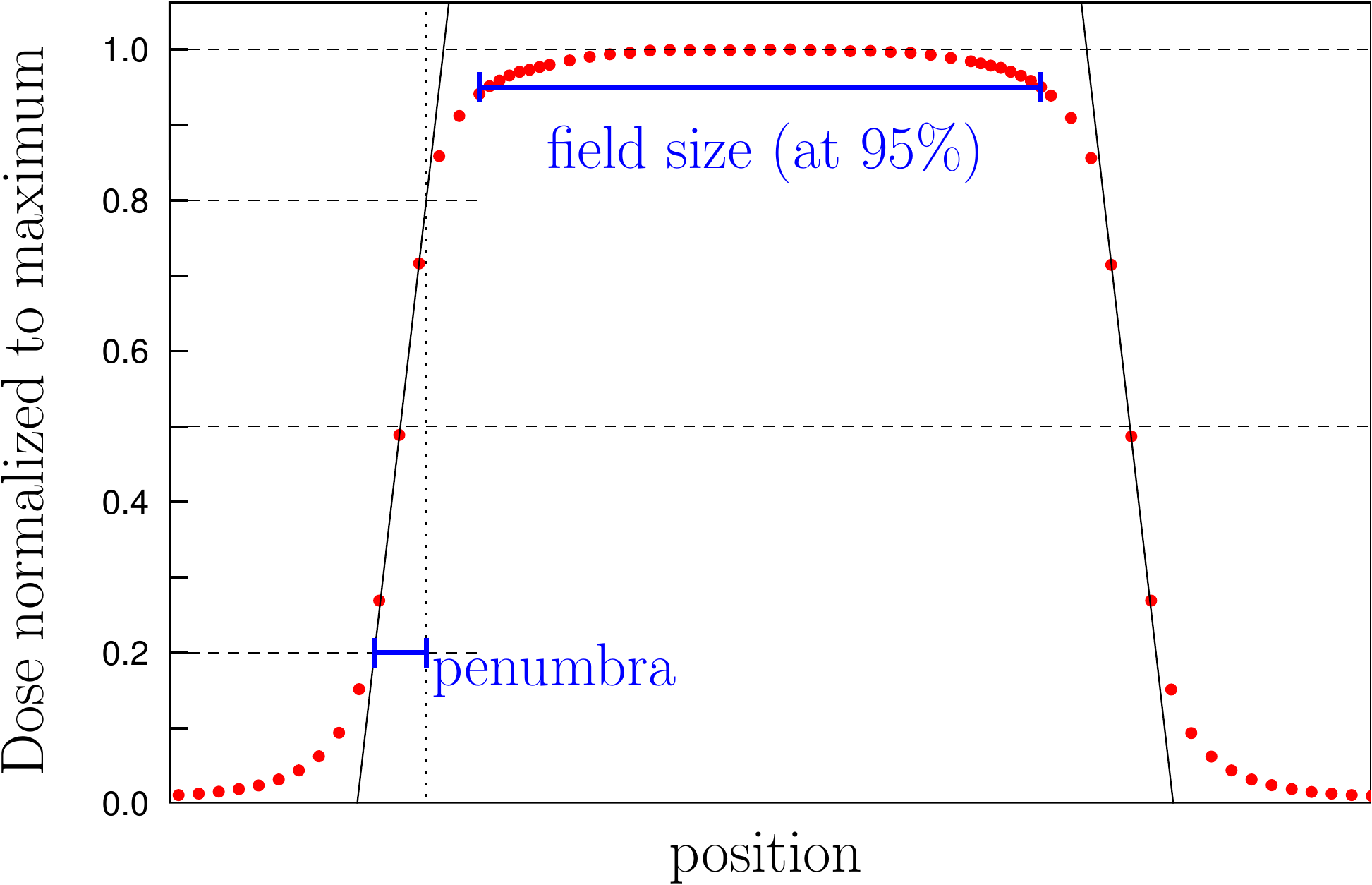}
\caption{Scheme indicating the definitions of the field size (at 95\% of the maximum dose) and the penumbra used in the present work.
\label{fig:defi}
}
\end{figure}

In order to analyze the various profiles calculated in the simulations, we have determined the field size and the penumbra of each one. Figure \ref{fig:defi} sketches the definitions of both quantities.

The field size is defined as the distance between the two positions having 95\% of the maximum dose of the profile. The level of 95\% is that required in clinics to fix the treatments. We require from the proposed collimator intended to substitute the current collimator without brass insert to maintain at least 95\% of the maximum dose along 2.4~cm of a centered lateral profile in the $x$ direction (corresponding to the inferior-superior direction of the eyeball) measured at a depth of 2~cm in a water phantom.

To obtain the penumbrae, the following procedure was carried out. First we fitted a straight line to the data around the 50\% level of the maximum dose. The penumbra was defined as the distance between the positions having 20\% and 80\% of the maximum dose, measured on the fitted line. We checked that the penumbra was the same on both sides of the $x$ profiles.

\section{Results and discussion}

For the various modified geometries analyzed, depth dose and lateral profiles are compared to those found with the G$_0$ geometry. The results obtained with this geometry are shown with solid circles while those corresponding to the modified geometries are plotted as colored curves. In each case, the differences between the doses obtained with the modified geometries and the original one are also shown as dotted curves. The depth dose distributions were obtained at $x = 0$ and $y = 1.3$~cm. The $x$ and $y$ lateral profiles have been obtained at $y = 1.3$~cm and $x = 0$, respectively, and for depths $z = 1.5$ (red), 3.0 (green) and 5.0 (blue)~cm.

\begin{figure}[!t]
\centering
\includegraphics[width=8cm]{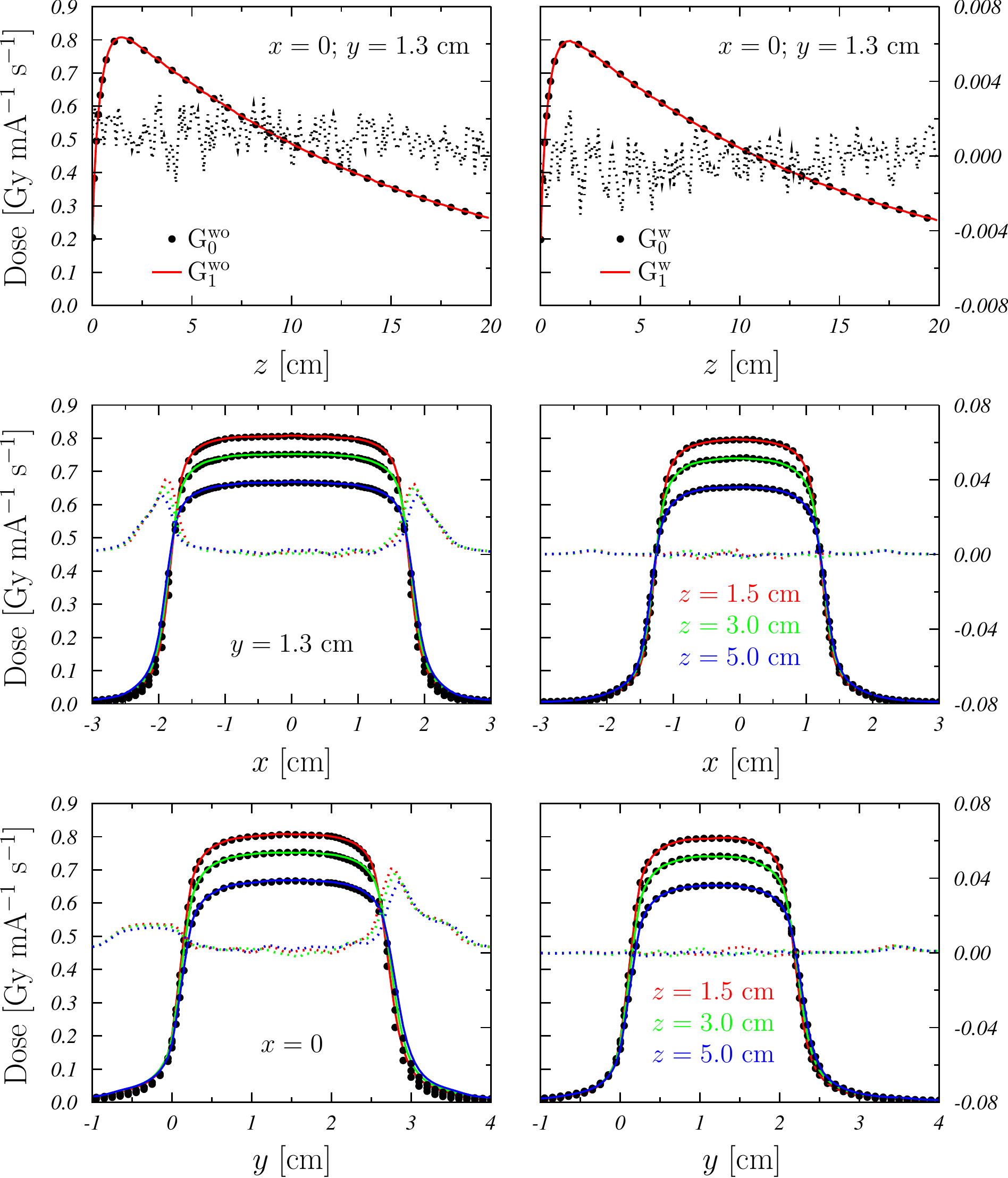}
\caption{Depth doses (upper panels) and $x$ (medium panels) and $y$ (lower panels) lateral profiles obtained with the geometry G$_{1}$ without (left panels) and with (right panels) the brass insert, represented with curves, are compared to those found with the reference geometry G$_0$ (points). The dotted curves are the differences $D_{{\rm G}_1}-D_{{\rm G}_0}$ (right-ordinate axes). 
\label{fig:G1}
}
\end{figure}

In figure \ref{fig:G1} the results found with the geometry G$_{1}$ are shown. This geometry permits to investigate the role played by the primary Rb collimator (see figure~\ref{fig:Gmod}). The differences between the results obtained G$_{0}$ and those with G$_{1}$ are very small, independently of the fact that the brass insert is included or not. As it can be seen in the medium and lower right panels, if the brass insert is used, the elimination of the primary Rb collimator does not produce any effect. If the brass insert is absent, a slight modification of the lateral profiles occurs. The changes are limited to the penumbra regions where the dose found for G$_{1}$ is greater than that corresponding to G$_{0}$ . This increased dose is very small in the region where the eye lens is situated ($y < 0$) and a little bit larger in the posterior pole of the eye ($y\geq 3$ cm). In any case, these results indicate that the Rb primary collimator must be kept if the brass insert is not used, in order to maintain the dose distribution found with the G$_{0}$ geometry. 

\begin{figure}[!t]
\centering
\includegraphics[width=8cm]{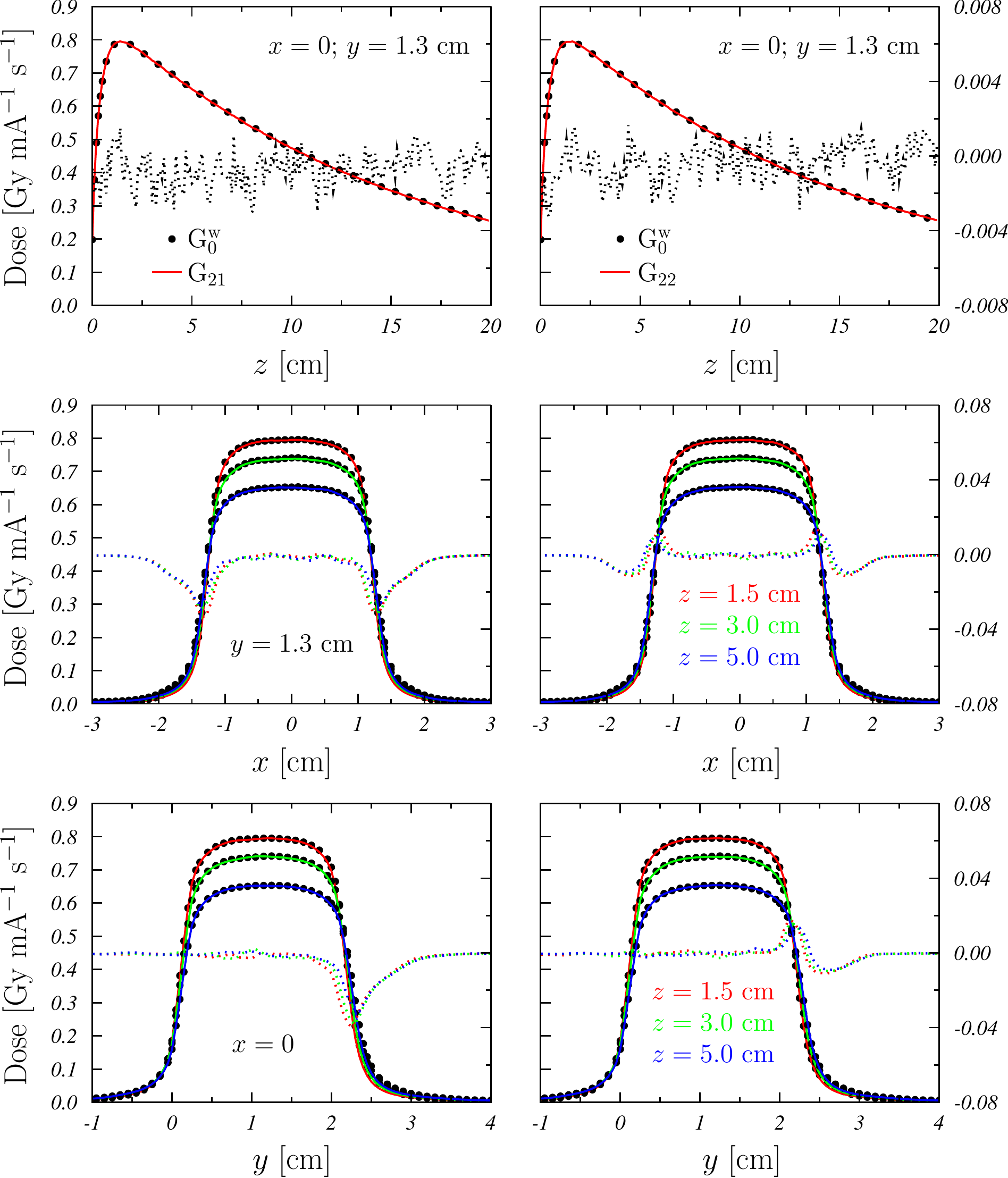}
\caption{Depth doses (upper panels) and $x$ (medium panels) and $y$ (lower panels) lateral profiles obtained with the geometries G$_{21}$ (left panels) and G$_{22}$ (right panels), represented with curves, are compared to those found with the reference geometry G$_0^{\rm w}$ (points). The dotted curves are the differences $D_{{\rm G}_{2\alpha}}-D_{{\rm G}_0^{\rm w}}$ (right-ordinate axes). 
\label{fig:G2}
}
\end{figure}

Figure \ref{fig:G2} shows the comparison of the G$_{21}$ (left panels) and G$_{22}$ (right panels) results to those found with the reference geometry. As indicated in figure \ref{fig:Gmod}, in the G$_{21}$ geometry, the material of the insert has been changed from brass to Cerrobend. This change produces a small variation of the penumbrae: the G$_{21}$ dose reduces with respect to the G$_0^{\rm w}$ one and this occurs only where the insert is present ($x \sim \pm1.4$~cm in the medium panel and $y \sim 2.2$~cm in the lower one). This means a dose reduction in the posterior pole of the eye without additional modifications of the dose absorbed at the eye lens. No relevant changes are observed in the depth dose distributions (upper panels). Also, as expected, there are no effects in the penumbrae of the $y$-lateral profiles (lower panels) at $y \sim 0$ (see figure \ref{fig:G0}) that remains as in G$_0^{\rm w}$. In geometry G$_{22}$, where the insert shape is modified to match the secondary Rb collimator (see figure \ref{fig:Gmod}), the effect seen in G$_{21}$ is largely compensated and the modifications in the penumbrae disappear to a large extent.

\begin{figure}[!t]
\centering
\includegraphics[width=8cm]{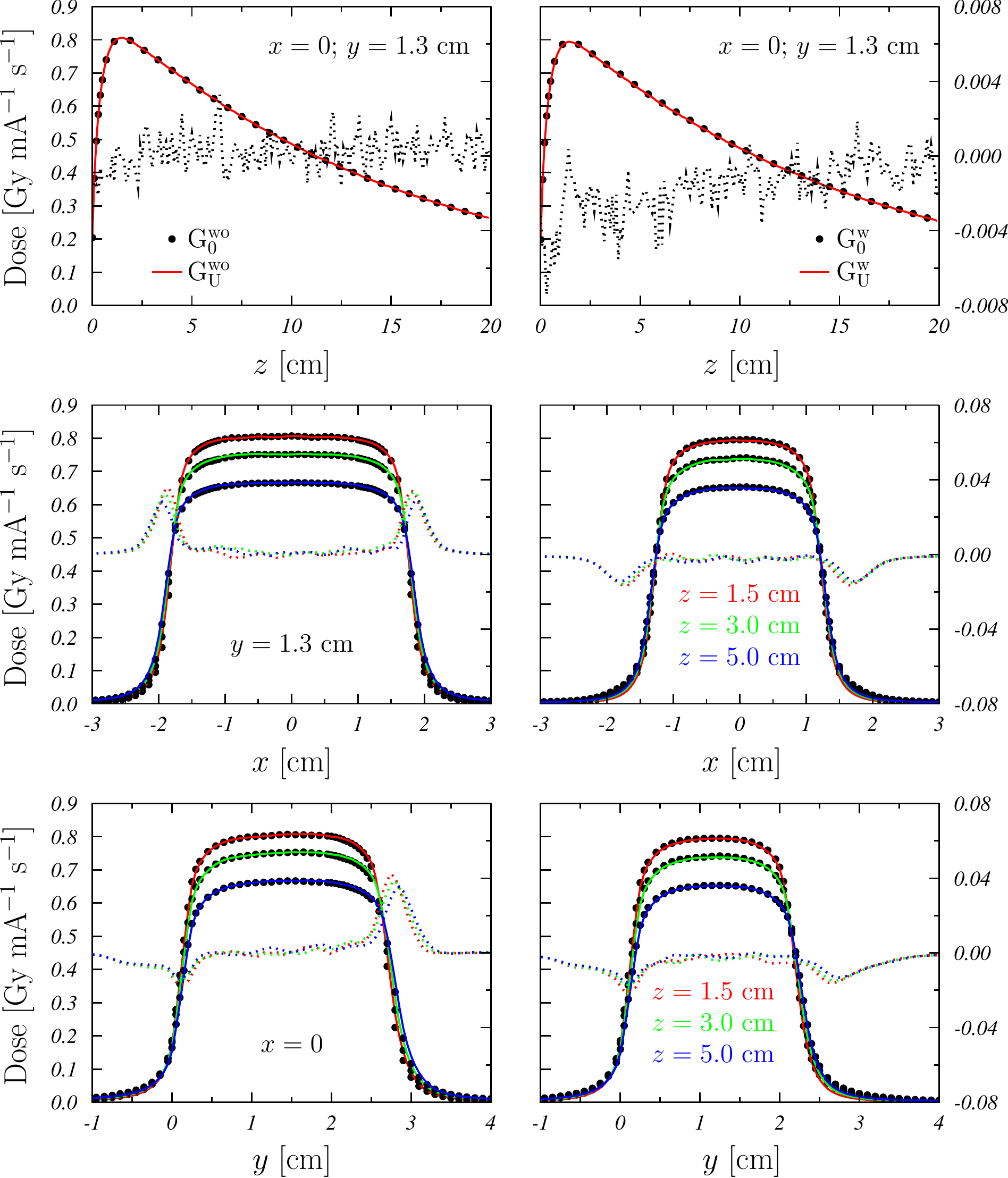}
\caption{Depth doses (upper panels), and $x$ (medium panels) and $y$ (lower panels) lateral profiles obtained with the geometry G$_{\rm U}$ without (left panels) and with (right panels) the brass insert, represented with curves, are compared to those found with the reference geometry G$_0$ (points).  The dotted curves are the differences $D_{{\rm G}_{\rm U}}-D_{{\rm G}_0}$ (right-ordinate axes). \label{fig:G3U}
}
\end{figure}

\begin{figure}[!th]
\centering
\includegraphics[width=8cm]{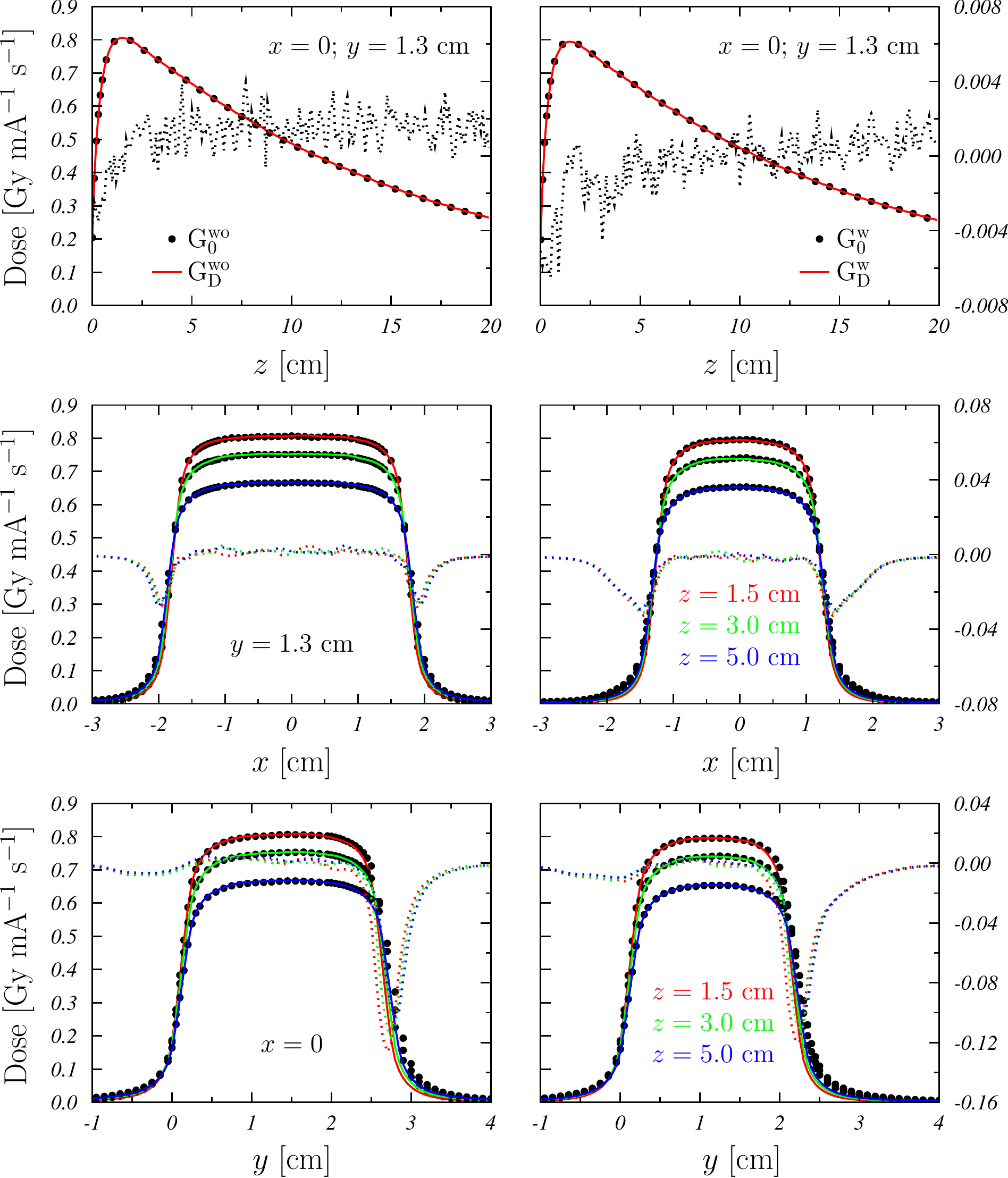}
\caption{Depth doses (upper panels), and $x$ (medium panels) and $y$ (lower panels) lateral profiles obtained with the geometry G$_{\rm D}$ without (left panels) and with (right panels) the brass insert, represented with curves, are compared to those found with the reference geometry G$_0$ (points). The dotted curves are the differences $D_{{\rm G}_{\rm D}}-D_{{\rm G}_0}$ (right-ordinate axes). \label{fig:G3D}
}
\end{figure}

The modifications produced on the absorbed dose distribution due to changes in the position of the jaws are shown in figures \ref{fig:G3U} and \ref{fig:G3D}. In the first one we have plotted the results corresponding to the geometry G$_{\rm U}$ in which the jaws were moved to circumscribe the field delimited by the upstream end of the secondary Rb collimator. The modifications on the depth dose distributions (upper panels) owing to the change of the jaws position is rather reduced independently of the presence of the brass insert. The modifications on the lateral profiles due to the aforementioned displacement of the jaws when the brass insert is present are also small and restricted to the penumbrae (medium and lower right panels). A slightly larger modification, also restricted to the penumbrae, is found when the insert is not included in the geometry. Therein, the doses found for G$_0$ are smaller than those obtained with G$_{\rm U}$ and a roughly 10\% increase of the dose at the posterior pole of the eye occurs.

\begin{figure}[!th]
\centering
\includegraphics[width=8cm]{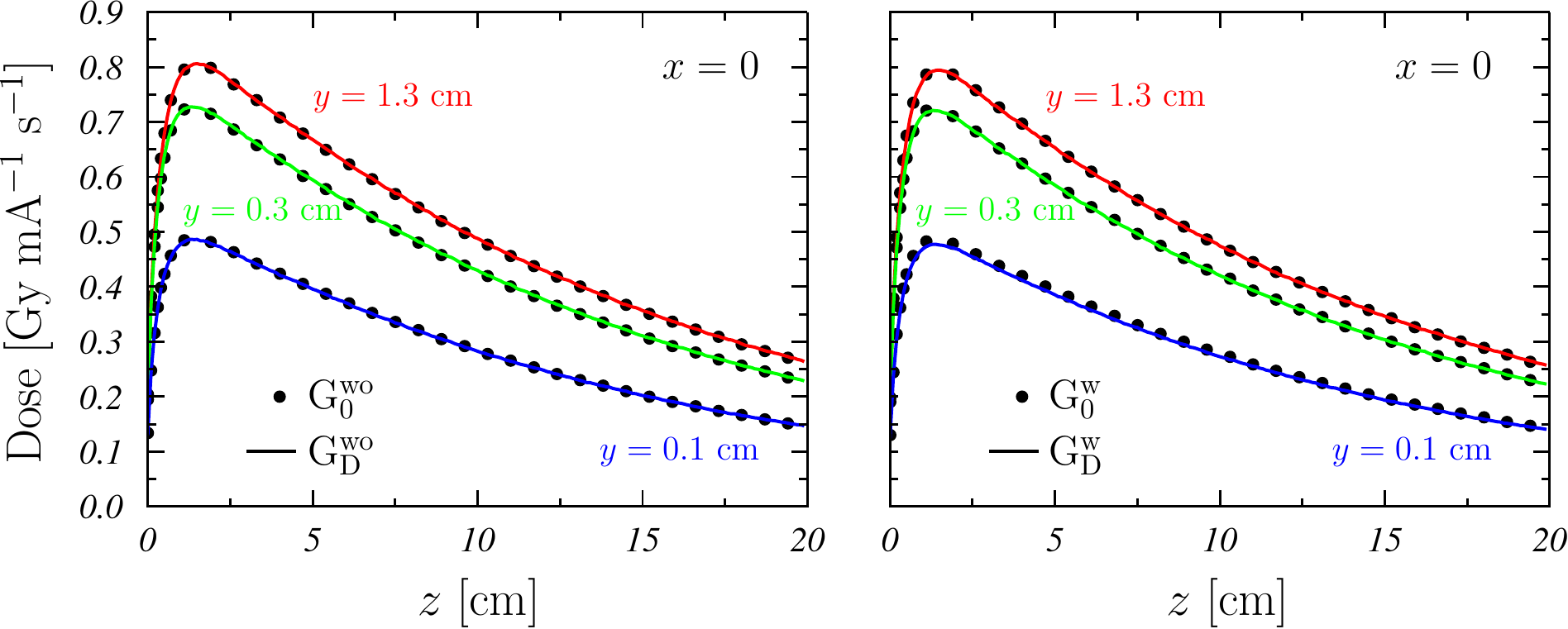}
\caption{Depth doses obtained at $x = 0$ and $y = 0.1$, $0.3$ and $1.3$~cm with the geometry G$_{\rm D}$ without (left panels) and with (right panels) the brass insert, represented with curves, are compared to those found with the reference geometry G$_0$ (points).
\label{fig:G3D-PDD}
}
\end{figure}

A different behavior is observed in the case of the G$_{\rm D}$ geometry (see figure \ref{fig:G3D}) in which the jaws were moved to circumscribe the field delimited by the downstream end of the secondary Rb collimator. The changes produced by the geometry modification are not relevant in the depth doses (upper panels). In G$_{\rm D}^{\rm wo}$, when the brass insert is not present, the effects of the modification in the jaws position on the lateral profiles (medium and lower left panel) change sign with respect to those found for G$_{\rm U}^{\rm wo}$ geometry (see figure \ref{fig:G3U}). These effects are of the same order for the $x$ profiles (medium panel) but appear to be largely increased in the case of the $y$ profiles. In fact, a reduction of $\sim 40$\% in the dose absorbed at the posterior pole is observed. Similar results are found when the brass insert is included (medium and lower right panels) but in this case the sign of the modifications does not change with respect to G$_{\rm U}^{\rm w}$.

In the case of the geometry G$_{\rm D}$, we have also investigated the effect that has on the depth dose profiles the fact of closening the jaws. In figure \ref{fig:G3D-PDD} we show these profiles, obtained for $x = 0$ and three different values of $y$: $1.3$~cm (which coincides with that shown in the upper panels of figure~\ref{fig:G3D}), $0.3$~cm and $0.1$~cm (this last very close to the end position of the field delimited by the jaws). We see that there is not a great effect due to the jaw situation. Only in the case in which the brass insert is present (right panel), for $y = 0.1$~cm, a small reduction in the maximum of the curves ($\sim 2$\%) is apparent.

Penumbrae are a relevant aspect of the dose profiles. Table~\ref{tab:penun} shows the values found for the different geometries studied at two depths in water, with and without the brass insert. In general, the increase of the depth produces slightly larger penumbrae. At $z = 3.0$~cm they are between 0.03 and 0.05~cm greater than those found at $z = 1.5$~cm. It is worth pointing out that the presence of the brass insert produces small modifications in the penumbra below 0.02~cm. The values obtained for the geometries including the brass insert are all statistically compatible with that obtained with the geometry G$_{0}^{\rm w}$. When the brass insert is absent greater differences occur at $z = 3.0$~cm among the values obtained for some of the geometries (see table~\ref{tab:penun}). 

\begin{table}[!b]
\caption{Penumbrae (in cm) of the $x$ profiles for two depths in water ($z=1.5$ and $3$~cm) for all geometries studied in the present work, including (w) or not (wo) the brass insert. Standard statistical uncertainties are 0.01~cm.
\label{tab:penun}}
\begin{center}
\begin{tabular}{cccccc}
\hline\hline
&\multicolumn{2}{c}{wo} &~~~&\multicolumn{2}{c}{w} \\
\cline{2-3}\cline{5-6}
$z$ (cm) & 1.5 & 3.0 && 1.5 & 3.0 \\\hline
G$_0$ & 0.31 & 0.34 && 0.32 & 0.36 \\
G$_1$ & 0.32 & 0.37 && 0.32 & 0.37 \\
G$_{21}$ & & && 0.31 & 0.35 \\
G$_{22}$ & & && 0.31 & 0.35 \\
G$_{\rm U}$ & 0.32 & 0.36 && 0.32 & 0.37 \\
G$_{\rm D}$ & 0.30 & 0.33 && 0.31	 & 0.35 \\ \hline
G$_{\rm opt}$ & 0.30 & 0.33 && & \\
\hline\hline
\end{tabular}
\end{center}
\end{table}

\begin{table}[!bh]
\caption{Field sizes (in cm) of the $x$ profiles for two depths in water ($z=1.5$ and 3~cm) for all the geometries studied in the present work, including (w) or not (wo) the brass insert. Standard statistical uncertainties are 0.04~cm.
\label{tab:field}}
\begin{center}
\begin{tabular}{cccccc}
\hline\hline
&\multicolumn{2}{c}{wo} &~~~&\multicolumn{2}{c}{w} \\
\cline{2-3}\cline{5-6}
$z$ (cm) & 1.5 & 3.0 && 1.5 & 3.0 \\\hline
G$_0$ & 2.75 & 2.70 && 1.65 & 1.65 \\
G$_1$ & 2.75 & 2.75 && 1.65 & 1.65 \\
G$_{21}$ & & && 1.65 & 1.65 \\
G$_{22}$ & & && 1.65 & 1.65 \\
G$_{\rm U}$ & 2.75 & 2.75 && 1.65 & 1.65 \\
G$_{\rm D}$ & 2.70 & 2.70 && 1.65 & 1.65 \\ \hline
G$_{\rm opt}$ & 2.45 & 2.45 && & \\
\hline\hline
\end{tabular}
\end{center}
\end{table}

An important point to be considered is the size of the radiation fields. Presently, the brass insert providing the smaller field is used to treat children younger than 6 months old. This field is, however, too small for older patients for whom a field of width 2.4~cm in the $x$-axis direction (inferior-superior direction in the eyeball) measured at 2~cm depth with 95\% of the maximum dose would suffice for an adequate treatment, even taking into account a safety margin of about 0.2~cm. Table~\ref{tab:field} summarizes the field widths obtained in our simulations at two of the depths analyzed, namely 1.5 and 3.0~cm. The field width along the $x$-axis does not change too much neither with the specific geometry considered nor with the depth in the water phantom. It is 1.65 cm when the brass insert is present and ranges from 2.70 to 2.75~cm without it. Our scope is to modify the geometry without the brass insert so to reduce its field width along the $x$-axis from 2.7~cm to 2.4~cm.

\begin{figure}[!t]
\centering
\includegraphics[width=8cm]{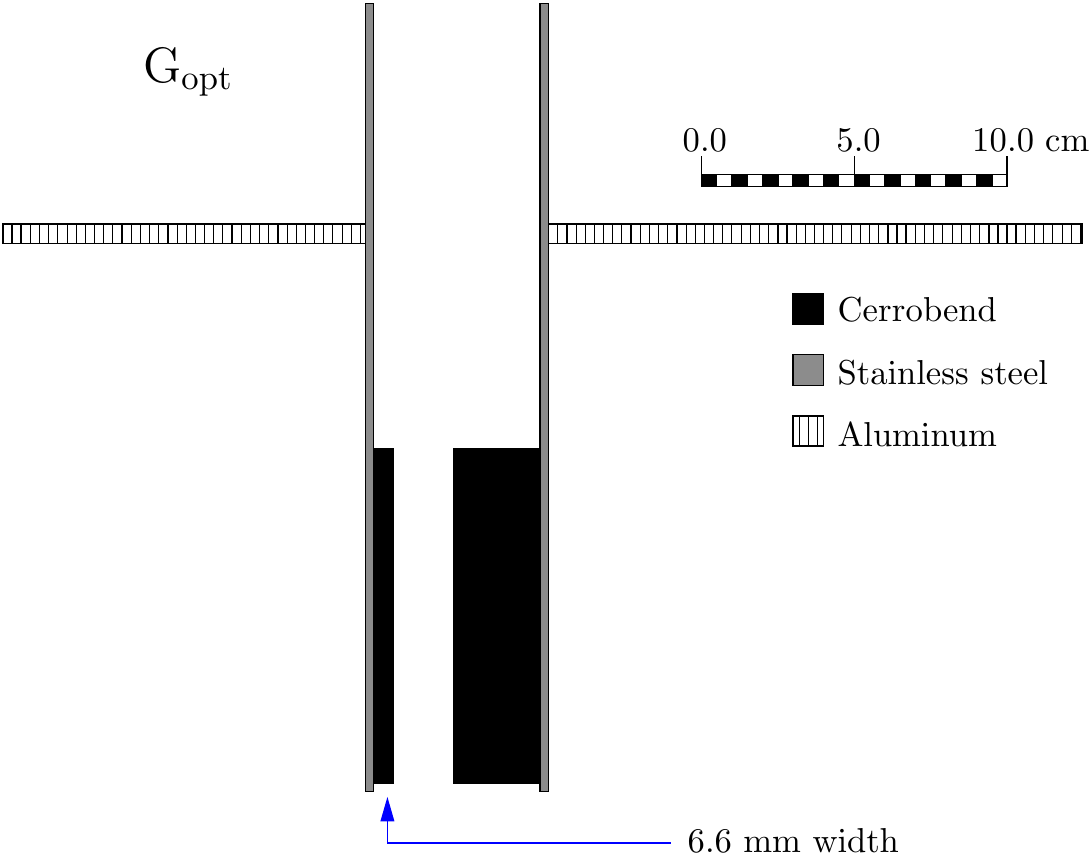}
\caption{Optimal geometry G$_{\rm opt}$ proposed in the present work. The primary collimator has been eliminated and the radius of the cylindrical surface limiting the curved part of the `D' has been reduced from 1.485~cm in G$_0^{\rm wo}$ to 1.365~cm. In addition, the jaws are closed as in the G$_{\rm D}$ geometry, so that they circumscribe the field delimited by the projections to the isocenter of the downstream end of the G$_{\rm opt}$ collimator. 
\label{fig:Geom-opt}
}
\end{figure}

\begin{figure}[!b]
\centering
\includegraphics[width=8cm]{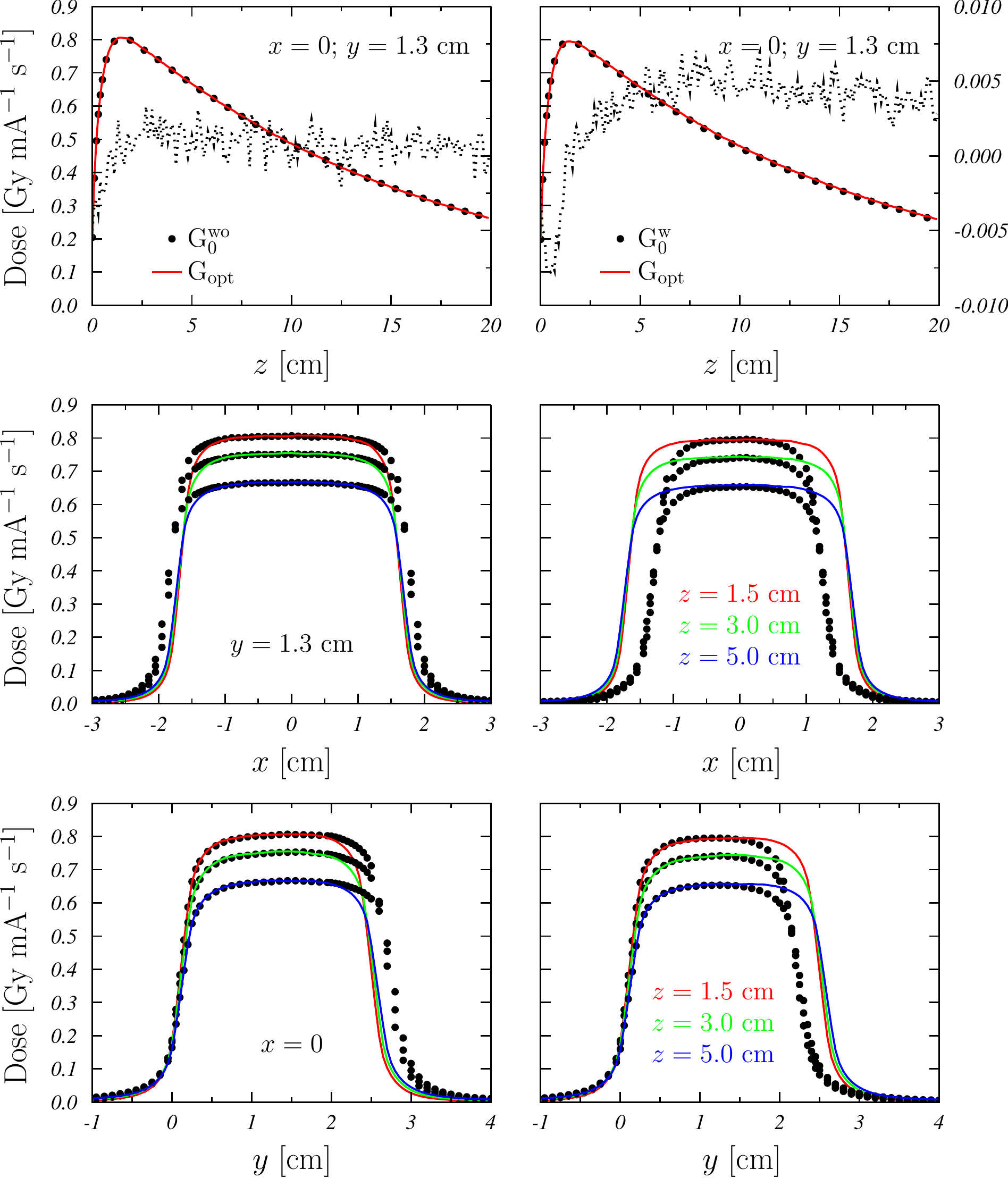}
\caption{Depth doses (upper panels) and $x$ (medium panels) and $y$ (lower panels) lateral profiles obtained with the geometry G$_{\rm opt}$, represented with curves, are compared to those found with the reference geometry G$_0$ without (left panels) and with (right panels) the brass insert (points). The dotted curves, plotted together with the depth dose profiles, are the differences $D_{{\rm G}_{\rm opt}}-D_{{\rm G}_0}$ (right-ordinate axes). 
\label{fig:Gopt}
}
\end{figure}

To ensure the therapeutic requirement about the field width we considered a new optimized geometry that we labeled G$_{\rm opt}$, shown in figure~\ref{fig:Geom-opt}. It includes the modifications of the geometries G$_1$ (in which the primary collimator was eliminated) and G$_{22}$ (in which the insert was modified to match the secondary Rb collimator). The radius of the cylindrical surface defining the curved part of the `D'-shaped collimator is 1.485~cm for the geometry G$_0^{\rm wo}$ and 1.085~cm for G$_0^{\rm w}$. For the proposed G$_{\rm opt}$ this radius is 1.365~cm. As in G$_{22}$ the collimator is made of Cerrobend (i.e., there is no brass). In addition, the jaws are positioned with the same scope followed in the G$_{\rm D}$ geometry, that is moving them in such a way that they circumscribe the field delimited by the projection to the isocenter of the downstream end of the secondary Rb collimator. 

In figure~\ref{fig:Gopt}, the results found with G$_{\rm opt}$ are compared to those obtained with G$_0^{\rm wo}$ (left panels) and G$_0^{\rm w}$ (right panels). To adequately perform the comparison shown in figure~\ref{fig:Gopt} the dose distribution obtained for the G$_{\rm opt}$ was multiplied by 1.024 and 1.009 when comparing with the G$_0^{\rm wo}$ and the G$_0^{\rm w}$ results, respectively. 

The upper panels of figure~\ref{fig:Gopt} show the depth dose distributions. The differences with the G$_0^{\rm wo}$ results are very small, except very close to the surface ($z\sim0$). However, these differences are more apparent in the case the brass insert is included. The expected reduction/increasing of the field size with respect to those defined by G$_0^{\rm wo}$ and G$_0^{\rm w}$ geometries is apparent in the medium panels. As we can see in table \ref{tab:field}, the field goes, respectively, from $\sim 2.7$ cm and $\sim 1.6$ cm to the $\sim 2.4$ cm in G$_{\rm opt}$. The large modification that this optimal geometry implies with respect to G$_0^{\rm w}$ is responsible for the differences observed in the case of the depth dose distributions. Also $y$-profiles (see lower panels) show the modification in the field size. 

Penumbrae (see table \ref{tab:penun}) do not change with respect to the other geometries and the variations with the depth in water are moderate. The field size does not change with $z$ and it maintains the required value of $\sim 2.4$~cm in all the range of depths analyzed.

\section{Conclusion}

In this work, the dosimetric impact produced by different geometry modifications introduced in the original design of the `D'-shaped collimation system currently in use for Rb treatments have been investigated. The changes affect the primary and secondary collimators of this system as well as the situation of the linac jaws delimiting the radiation field at the entrance of the Rb collimation system.

The various geometries analyzed have a rather small impact on the depth dose distributions at the beam center. The changes produced in both $x$ and $y$ lateral profiles are limited to the penumbrae and result to be relatively important only in the case of the G$_{\rm D}$ geometry which corresponds to a modification in the jaw position in order to delimit the radiation field that they define to that conformed by the downstream end of the secondary Rb collimator.

An optimal geometry G$_{\rm opt}$ has been proposed; it produces a field of width $2.4$~cm along $x$-axis, corresponding to the inferior-superior direction of the eyeball, at $2$~cm depth with 95\% of the maximum dose. With this new design an important reduction of the irradiated volume is obtained, without reducing the dose to the target volume of the patients that are currently treated with the G$_0^{\rm wo}$ collimator.


\acknowledgments{The work of PAM and AML has been supported in part by the Junta de
Andaluc\'{\i}a (FQM0220) and FEDER and the Spanish Ministerio de Ciencia e
Innovaci\'{o}n and FEDER (project FPA2009-14091-C02-02) and Ministerio de Econom\'{\i}a y Competitividad (project FPA2012-31993). LB is grateful to the Deutsche Forschungsgemeinschaft (project BR~4043/1-1).}

\end{document}